\documentclass[a4paper,fleqn,usenatbib]{mnras}

\usepackage{newtxtext,newtxmath}
\usepackage{multirow}
\usepackage[T1]{fontenc}
\usepackage{ae,aecompl}

\usepackage{graphicx}	% Including figure files
\usepackage{amsmath}	% Advanced maths commands
\title[Dust temperatures at high redshift]{A simple numerical experiment on the dust temperature bias
for Lyman break galaxies at $z\gtrsim 5$}

\author[Y. Y. Chen et al.]{
Yung Ying Chen,$^{1}$\thanks{E-mail: abbey000508@gmail.com} Hiroyuki Hirashita,$^{1}$ Wei-Hao Wang$^1$ and Naomasa Nakai$^2$
\\
$^{1}$Institute of Astronomy and Astrophysics, Academia Sinica,
Astronomy-Mathematics Building, No.\ 1, Section 4,
Roosevelt Road, Taipei 10617, Taiwan\\
$^2$School of Science and Technology, Kwansei Gakuin University, 2-1 Gakuen, Sanda, Hyogo 669-1337, Japan
}

\date{Accepted 2021 October 25. Received 2021 October 6; in original form 2021 July 26}

\pubyear{2021}

\begin{document}
\label{firstpage}
\pagerange{\pageref{firstpage}--\pageref{lastpage}}
\maketitle

% Abstract of the paper
\begin{abstract}
%%Using the correct dust temperature is a key to derive the total infrared (IR) luminosity and the dust mass of a galaxy from limited photometric data points. 
Some studies suggest that the dust temperatures ($T_\mathrm{d}$)
in high-redshift ($z\gtrsim 5$) Lyman break galaxies (LBGs) are high.
However, possible observational bias in $T_\mathrm{d}$ is yet to be understood.
Thus, we perform a simple test using random realizations of LBGs with various stellar masses, dust temperatures, and dust-to-stellar mass ratios, and examine how the sample detected by ALMA is biased in terms of $T_\mathrm{d}$.
We show that ALMA tends to miss high-$T_\mathrm{d}$ objects even at total dust luminosity $L_\mathrm{IR}>10^{11}~\mathrm{L}_{\sun}$.
%%, that is, the detection probability for low-$T_\mathrm{d}$ object are higher than those with high-$T_\mathrm{d}$ ones.
%%Thus, high-$T_\mathrm{d}$ objects tend to be missed by the currently available sensitivity achieved by ALMA.
%%Compared to 850 and 1200$\micron$, the 450 $\micron$ samples are less biased, but the detection rate is much less.
LBGs are, however, basically selected by the stellar UV luminosity.
The dust-temperature bias in a UV-selected sample is complicated because of the competing effects between high $T_\mathrm{d}$ and low dust abundance. For ALMA Band 6,
%%often used for high-$z$ dust observations,
there is no tendency of high-$T_\mathrm{d}$ LBGs being more easily detected in our experiment. Thus, we suggest that the observed trend of high $T_\mathrm{d}$ in $z\gtrsim 5$ LBGs is real. %%The low detection rate of high-$z$ LGBs is also consistent with the adopted dust abundances.
We also propose that the 450 $\micron$ band is useful in further clarifying the dust temperatures.
To overcome the current shallowness of 450 $\micron$ observations, we
examine a future Antarctic 30-m class telescope with a suitable atmospheric condition for wavelengths $\lesssim 450~\micron$, where the detection is not confusion-limited.
We find that, with this telescope, an $L_\mathrm{IR}$-selected sample with $\log(L_\mathrm{IR}/\mathrm{L}_{\sun})>11$ is constructed for $z\gtrsim 5$, and
detection in the intermediate-$M_\star$ (stellar mass) range [$9<\log (M_\star /M_{\sun})<9.5$] is much improved, especially at high $T_\mathrm{d}$.
%%It also proves that there is a bias towards high $T_\mathrm{d}$ for a UV-selected sample followed up at 450 $\micron$
%%because of the capability of detecting high-$T_\mathrm{d}$, low dust abundance objects.
\end{abstract}


%%%%%%%%%%%%%%%%% BODY OF PAPER %%%%%%%%%%%%%%%%%%

%%\usepackage[utf8]{inputenc}
\section{introduction }
\label{sec:introduction}
Dust grains are widespread in the interstellar medium (ISM) that is the ingredient of star formation. Dust also absorbs and scatters the radiation from formed stars in the ISM, and reprocesses it in the infrared (IR)--submillimetre (submm) \citep[e.g.][]{Buat:1996a,Calzetti:2000a}. These radiative processes of dust are important in the following two aspects. First, to estimate the star formation rate in a galaxy using the stellar light, correction for dust extinction is crucial \citep[e.g.][]{Steidel:1999a,Inoue:2000a}. Secondly, dust largely affects the spectral energy distributions (SEDs) of galaxies \citep[e.g.][]{Silva:1998a,Takagi:2003a,Takeuchi:2005}. Since SEDs are used to extract various information (stellar mass, age, etc.; e.g. \citealt{da-Cunha:2008a,Boquien:2019a}), it is fundamentally important to appropriately consider dust extinction and emission. Dust also affects the physical and chemical states of the ISM. Dust surfaces provide reaction sites for forming some molecular species \citep[e.g.][]{Chen:2018a}, especially molecular hydrogen \citep[e.g.][]{Gould:1963,Cazaux:2004a}, leading to the formation of molecule-rich environments in galaxies \citep[e.g.][]{Hirashita:2002,Yamasawa:2011}. In the star formation process, dust cooling induces fragmentation of molecular clouds and shapes the stellar initial mass function (IMF) \citep[e.g.][]{Whitwirth:1998a,Schneider:2006,Omukai:2005a,Larson:2005a}.
Because of the above important roles of dust, it is crucial to clarify the origin and evolution of dust in the Universe.

Observing the high-redshift Universe is useful to know how galaxies are enriched with dust
in the early phase of their evolution.
The current frontier of dust observation lies at $z\gtrsim 5$ \citep[e.g.][]{Capak2015a,Casey:2018a,Burgarella:2020a,Zavala:2021a}, where $z$ is the redshift; thus, in this paper,
high redshift indicates $z\gtrsim 5$.
With the Atacama Large Millimetre/submillimetre Array (ALMA), dust emission from high-redshift galaxies has become accessible \citep[e.g.][]{Dayal:2018a}. In particular, dust emission has been detected for a ‘typical' population of high-redshift galaxies, Lyman break galaxies (LBGs), even at $z>7$ \citep{Watson:2015a,Laporte:2017a,Tamura:2019a,Hashimoto:2019a,Schouws:2021a}.
However, most of the LBGs at such high redsdhift remain undetected even with ALMA \citep[e.g.][]{Riechers:2014,Bouwens:2016a,Fudamoto:2020a}.

The dominant source of dust at high redshift is still being debated \citep[e.g.][]{Lesniewska:2019a}.
Dust grains condense in stellar ejecta: Supernovae (SNe) are expected to be the first dust sources in the Universe because their progenitors have short lifetimes \citep[e.g.][]{Todini:2001a,Nozawa:2003a}. However, a part of the dust formed in SNe could be destroyed in the shocked region before being injected into the ISM \citep[e.g.][]{Bianchi:2007a,Nozawa:2007a}. Thus, it is not obvious if SNe produce a sufficient amount of dust in the early stage of galaxy evolution. To ‘supplement' the dust abundance, it has been suggested that dust growth by the accretion of gas-phase metals in the dense ISM dominates the increase of dust mass in some high-redshift galaxies \citep[e.g.][]{Mancini:2015a,Wang:2017a,Liu:2019a}.

To observationally reveal the dust sources at high redshift, it is important to correctly estimate the dust mass.
The estimate of dust temperature is particularly important in deriving the dust mass in
galaxies. Even if we fix the dust mass absorption coefficient (which is generally uncertain),
the uncertainty (or the lack of knowledge) in the dust temperature highly nonlinearly affects the
dust mass estimate. Some studies estimated the dust temperatures in high-redshift LBGs
basically from rest-frame far-IR (FIR) multi-wavelength data.
\citet{Knudsen:2017}, using the detection of a LBG at $z=7.5$ (A1689zD1; first detected with ALMA by \citealt{Watson:2015a}) in ALMA Band 6 and Band 7, obtained a dust temperature of 35--45 K, which is higher than the typical dust temperatures in the local spiral galaxies ($\sim 20$--25 K; e.g.\ \citealt{Draine:2007a,Skibba:2011a}). 
\citet{Inoue:2020a} and \citet{Bakx:2021a} further added a detection in Band 8 and 9, respectively, and confirmed the high dust temperature.
\citet{Burgarella:2020a} compiled detected LBGs at various $z(>5)$ to trace SEDs at different restframe FIR wavelengths. In this way, they derived typical dust temperatures of 40--70 K, higher than those in local spiral galaxies.
\citet{Bakx:2020a} obtained a dust temperature of $>80$ K for a LBG at $z=8.31$.
\citet{Faisst:2020a} observed four LBGs at $z\sim 5.5$ and derived high dust temperatures of 30--43 K.
These high dust temperatures at high redshift imply that adopting a typical dust
temperature in nearby spiral galaxies systematically overestimates the dust mass at high redshift.

%%In general, observational estimates of dust mass are uncertain,
%%especially at high redshift, because the flux measurements are uncertain and the dust material properties are unknown. More importantly, the uncertainty in the dust temperature gives rise to a large uncertainty in the obtained dust mass. The dust temperature is particularly uncertain for high-redshift LBGs because of the difficulty in obtaining a precise flux. Moreover, the determination of dust temperature requires observations at two wavelengths, one of which should ideally be near the peak of dust emission SED. Because of the limited atmospheric windows for ground-based telescopes and limited spatial resolutions (i.e.\ severe galaxy confusion) for space far-IR telescopes, the determination of dust temperature for high-redshift galaxies is usually challenging.

The dust temperatures also give a clue to the physical condition in
high-redshift galaxies.
A trend of higher dust temperatures at higher redshift is also seen at $z\lesssim 4$ \citep{Bethermin:2015a,Schreiber:2018a}, and could be related to higher star formation efficiencies \citep{Magnelli:2014a}.
There are other possible theoretical reasons for high dust temperatures in high-$z$ LBGs. For example,
if high-redshift galaxies host compact star-forming regions or high surface densities of star formation rate, the dust temperature is expected to be higher \citep{Ferrara2017a,Ma:2019a,Sommogigo:2020a}.

The above understanding of the dust temperatures in high-redshift galaxies is far from complete and is perhaps biased.
%%This is because the statistical properties of dust temperature at high redshift are difficult to clarify. Observationally, the sample size of LGBs detected by ALMA is still limited at $z\gtrsim 5$
%%with most LGBs not detected by ALMA \citep[e.g.][]{Bouwens:2016a}.
It is difficult to obtain fluxes from high-redshift LBGs in multiple ALMA bands precisely enough to determine a well constrained
dust temperature. The detectability is also
affected by the dust temperature; in particular, if the dust temperature is high as indicated above,
ALMA submm--millimetre bands may miss the peak of dust emission located at shoter wavelengths
\citep[e.g.][]{Hirashita:2017a}. If the dust temperature is low, the emission is inefficient so that the dust is faint in the ALMA bands. These effects of dust temperature could hamper our unbiased understanding of dust evolution in high-redshift galaxies.

In principle, the dust temperature bias can be addressed if we theoretically
predict the dust temperature distribution of high-redshift LBGs. However, predicting the statistical properties of
dust temperature is not easy because of the following issues. The first problem is low
spatial resolution.
Although some cosmological simulations \citep[e.g.][]{Springel:2003a} successfully included dust evolution \citep[e.g.][]{McKinnon:2017a,Aoyama:2018a,Hou:2019a,Graziani:2020a} and predicted dust temperatures \citep[e.g.][]{Aoyama:2019a}, simulations on galactic or larger scales generally have a low spatial resolution. Because of this limitation, an intense radiation field from compact star-forming regions, which could be important for explaining the high dust temperatures at high redshift, is difficult to investigate.
Spatially unresolved treatments such as semi-analytic models
\citep{Valiante:2011a,de-Bennassuti:2014a,Popping:2017a,Ginolfi:2018a}
and post-processing models \citep{Mancini:2016a,Huang:2021a} have difficulty in predicting
the dust temperature.
The second is a small sample size.
Some zoom-in simulations succeeded in investigating the details of dust distribution \citep{Yajima:2015a,McKinnon:2016a,Gjergo:2018a,Granato:2021a} and dust temperatures \citep[e.g.][]{Ma:2019a,Liang:2019a,DiMascia:2021a}, but the conclusion may rely on the zoomed particular objects.

In this paper, we aim at clarifying if there is any bias for the dust temperature
in a sample of high-redshift ($z\gtrsim 5$) LBGs observed by ALMA. Since theoretical methods (i.e.\ simulations
and semi-analytic models) have limitation as mentioned above, we
take a different, simple approach based on random realizations of LBGs; that is, we construct a virtual big sample of LBGs that enables us to examine the statistical properties of the dust temperatures.
The realizations are based on random sampling of some fundamental observational
quantities whose ranges are constrained empirically by observed LBGs at $z\gtrsim 5$. In this way, a big sample is easy to generate. Based on this virtual sample, we examine how the detected LBGs are biased in terms of the dust temperature. 
%%We construct an `experimental' sample of LBGs that enables us to survey
%%the relevant parameter space, and
%%we virtually observe the sample LBGs to judge if each of them is detectable or not.
%%This eventually enables us to examine if the dust temperatures of detected objects are biased or not.
The bias clarified through this approach will serve to judge if the observed high dust temperatures at high redshift reflect a real trend or an observational selection effect.
%%Since high-redshift LBGs are selected at rest-frame UV wavelengths, we also refer to LBGs as UV-selected galaxies.

In addition, we investigate a possibility of improving the dust temperature estimate.
Besides the often used bands at 850 and 1200~$\micron$ (Band~7 and 6, respectively), we add 450~$\micron$ (Band 9), which is near to the SED peak of galaxies at $z\gtrsim 5$ or even on Wien's side depending on the dust temperature \citep[e.g.][]{Bakx:2021a}.
The usefulness of the 450 $\micron$ band is demonstrated at lower redshifts ($z<5$).
\citet{Casey:2013a} showed that the dust temperatures of 850 and 450 $\micron$ samples in their survey using the SCUBA-2 instrument on the James Clerk Maxwell Telescope (JCMT) are different: The 450 $\micron$ band tends to detect galaxies with higher dust temperatures.
Recently, the SCUBA-2 Ultra Deep Imaging East Asian Observatory Survey (STUDIES) has been conducted at 450 $\micron$,
starting to detect galaxies at the knee of the IR luminosity function
up to $z\sim 3$ \citep{Wang:2017b}. Optical counterpart identifications and multi-wavelength SED fitting were successfully performed up to $z\sim 4$ \citep{Lim:2020a,Dudzevi:2021a}.
It is highly probable that the 450 $\micron$ band is also useful at $z\gtrsim 5$; thus, we discuss the detectability at 450 $\micron$ in this paper.

As we will show later, the worse sensitivity at 450$\micron$ is worse than at 850 and 1200 $\micron$
limits the dust temperature studies at $z\gtrsim 5$.
For a further improvement of 450 $\micron$ observations,
we consider a future Antarctic large single-dish telescope as a representative plan, and
discuss if such a future telescope contributes to a further understanding of the dust temperatures in high-redshift LBGs.
Observations at such a short submm wavelength, or at a nearly tera-hertz (THz) frequency, require a low water vapour content in the atmosphere, and are best carried out from Antarctic sites and Greenland \citep[e.g.][]{Hirashita:2016a,Matsushita:2017a}.
A similar scientific goal could also be achieved by future $>$30-m-class submm telescopes such as the 
Atacama Large-Aperture Submmillimetre/millimetre Telescope \citep{Klaassen:2020aa} and the Large Submillimetre Telescope \citep{Kawabe:2016aa}.

%%However, the statistical properties of dust temperatures
%%beyond $z\sim 4$ are still unclear. Future sensitive telescopes will
%%help us extend the above studies to $z\gtrsim 5$.

This paper is organized as follows. We explain the method for generating a LBG sample in Section~\ref{sec:model}. We show the results in Section~\ref{sec:result}. We discuss some further issues, especially possible improvement using a future telescope in Section \ref{sec:discussion}. Section \ref{sec:conclusion} concludes this paper. We adopt the following cosmological parameters: cosmological constant parameter $\Omega_\Lambda=0.7$, total matter density parameter $\Omega_\mathrm{M}=0.3$, and Hubble constant $H_{0}= 70$ km s$^{-1}$ Mpc$^{-1}$.

\section{Model}\label{sec:model}
We use a simple method based on random sampling of the relevant parameter ranges for basic quantities
that characterize the dust emission in high-redshift ($z\geq 5$) LBGs.
We focus on LBGs as a representative population for high-redshift galaxies, and exclude extreme populations such as submm galaxies (SMGs) and quasars.
We first describe our assumptions.
Next, we discuss how the basic quantities are related to the dust emission
luminosity and the observed flux. Finally, for the purpose of parameter surveys, we generate random realizations of LBGs, which are
virtually observed to discuss possible temperature biases in detected objects.

\subsection{Assumptions}
\label{assumptions}
We assume that the dust emission from a LBG is characterized by stellar mass $M_\star$,
dust-to-stellar mass ratio $\mathcal{D}_\star$, and dust temperature $T_\mathrm{d}$. LBGs at $z\gtrsim 5$ are mostly selected by rest-frame UV flux, which is well correlated with the stellar mass \citep{Schaerer:2015a} (see also Section \ref{subsec:model_LIR}). Thus, we assume that the stellar mass is one of the most fundamental quantities.
Indeed, \citet{Burgarella:2020a} normalized both star formation rate and dust mass by the stellar mass,
and found a meaningful relation between these two quantities. This supports our idea of using $M_\star$ for the overall scaling
factor. Moreover, we also expect that the dust enrichment, strongly related to metal enrichment, proceeds along with the buildup of
stellar mass (or the metal enrichment associated with star formation; e.g.\ \citealt{Tinsley:1980a}). To obtain the dust mass, $M_\mathrm{d}$, we use the dust-to-stellar mass ratio
($\mathcal{D}_\star$; referred to as the specific dust mass in \citealt{Burgarella:2020a}) for the second parameter. Using this quantity,
the dust mass $M_\mathrm{d}$ is given by
$M_\mathrm{d}=\mathcal{D}_\star M_\star$.
Finally, with a given dust mass (and dust mass absorption coefficient), the total dust luminosity is determined by the dust temperature.
Thus, we use $T_\mathrm{d}$ for the third parameter. Since $T_\mathrm{d}$ is used to estimate the dust emission luminosity,
$T_\mathrm{d}$ is interpreted as luminosity-weighted dust temperature
\citep[e.g.][]{Liang:2019a}.

As mentioned above, we assume that the dust heating sources (i.e.\ stars) are mostly traced by rest-frame UV observations.
This reflects the fact that high-redshift LBGs are first sampled by their rest-frame UV emission.
Because of this selection, we do not treat highly embedded star formation
activities as seen in extreme starbursts (such as SMGs). This assumption is
equivalent to the hypothesis that the total IR luminosity $L_\mathrm{IR}$ is not
much higher than the UV luminosity $L_\mathrm{UV}$. The ratio $L_\mathrm{IR}/L_\mathrm{UV}$ is
referred to as the infrared excess (IRX). Indeed, $\mathrm{IRX}\lesssim 10$ for high-$z$ LBGs
\citep[][]{Bouwens:2016a,Burgarella:2020a}.
(We discuss the value of IRX further in  Section \ref{subsec:parameter}.)
Thus, we adopt the set of parameters $(M_\star,\,\mathcal{D}_\star,\, T_\mathrm{d})$
under the constraint that IRX does not exceed a certain value ($\sim 1$--10).

%%As mentioned in Section \ref{subsec:IRXeffect}, precisely speaking,
%%$\mathcal{D}_\star$ and $T_\mathrm{d}$ are, respectively, the abundance and temperature of the dust %%heated by the stars.\
%%Thus, $\mathcal{D}_\star$ and $T_\mathrm{d}$ may not represent the entire dust population in a LBG.
%%Since submm observations can, in general, only trace such a heated dust component, the quantities we give to each LBG are chosen from the actually observed ranges (Section \ref{subsec:parameter}).

At high redshift, the effect of the CMB heating on the dust temperature may not be negligible. Thus, $T_\mathrm{d}$ is not totally free because it cannot drop below the CMB temperature. To include this effect, we first input the `virtual' dust temperature, $T_\mathrm{d}^0$, which would be realized if there is no heating from the CMB, and then correct it for the CMB heating. We treat $T_\mathrm{d}^0$ as a free input parameter. The dust temperature after correcting for the CMB effect is obtained by \citep{da-Chnha:2013a}
\begin{align}
T_\mathrm{d}=\left\{(T_\mathrm{d}^0)^{4+\beta_\mathrm{IR}}+(T_\mathrm{CMB}^0)^{4+\beta_\mathrm{IR}}\left[
(1+z)^{4+\beta_\mathrm{IR}} -1\right]\right\}^\frac{1}{4+\beta_\mathrm{IR}} ,\label{eq:CMB}
\end{align}
where $\beta_\mathrm{IR}$ is the dust emissivity index (given in equation \ref{eq:kappa})
and $T_\mathrm{CMB}^0$ is the CMB temperature at $z=0$ (2.73 K).

\subsection{Calculation of the dust luminosity and flux}\label{subsec:model_LIR}
We assume that the dust emission follows the so-called modified blackbody spectrum
with a single dust temperature.
Thus, the monochromatic luminosity of the dust emission at rest-frame frequency $\nu$ (denoted as
$L_\nu$) is estimated by the following equation \citep[e.g.][]{Dayal:2010a}:
\begin{equation}
L_{\nu}=4\pi \kappa_{\nu} M_\mathrm{d} B_{\nu} (T_\mathrm{d}),
\label{eq:L_IR}
\end{equation}
where $\kappa_{\nu}$ is the mass absorption coefficient at frequency $\nu$, and $B_{\nu}(T_\mathrm{d})$
is the Plank function at frequency $\nu$ and temperature $T_\mathrm{d}$. The mass absorption
coefficient is estimated by assuming a power-law form as \citep[e.g.][]{Hirashita:2014a}
\begin{equation}
\kappa_{\nu}=\kappa_{158}\left(\frac{\nu}{\nu_{158}}\right)^{\beta_\mathrm{IR}},
\label{eq:kappa}
\end{equation}
where $\kappa_{158}$ is the value at a wavelength ($\lambda$) of 158 $\micron$,
$\nu_{158}$ is the frequency corresponding to $\lambda =158~\micron$, and
$\beta_\mathrm{IR}$ is the dust emissivity index. The choice of the wavelength for normalization
is arbitrary.
For the dust species, we adopt the following often adopted materials: silicate, graphite, and amorphous carbon (AC).
We adopt the values of $\kappa_{158}$ and $\beta_\mathrm{IR}$ for each species as listed in Table \ref{tab:param}.
We use graphite unless otherwise stated, since it has an intermediate mass absorption coefficient.
We discuss the other species in Section \ref{subsec:species}.
%%Since we consider CMB effect in parameter setting for temperature (Section \ref{subsec:parameter}),
The observed flux at frequency $\nu$ (denoted as $f_\nu$) is estimated as  \citep{da-Chnha:2013a}
\begin{equation}
f_{\nu}=\frac{(1+z)}{{d_L}^2}
\kappa_{\nu(1+z)}\,M_\mathrm{d}\left[ B_{\nu(1+z)}(T_\mathrm{d})-B_{\nu(1+z)}(T_\mathrm{CMB})\right] ,
\end{equation}
where $T_\mathrm{CMB}=T_\mathrm{CMB}^0(1+z)$ is the CMB temeprature at redshift $z$ and $d_L$ is the luminosity distance given by e.g.\ \citet{Carroll:1992a}.

\begin{table}
\caption{Dust species}\label{tab:param}
\begin{center}
\begin{tabular}{lccc}
\hline
Species & $\kappa_{158}$ & $\beta_\mathrm{IR}$ & $C_\mathrm{IR}$ \\
 & (cm$^2$ g$^{-1}$) & & (see text) \\
\hline
Silicate & 13.2 & 2 &  $3.5\times 10^{-6}$\\ %%3.522e-06 \\
Graphite & 20.9 & 2 &  $5.6\times 10^{-6}$\\ %%5.577e-06 \\
AC & 28.4 & 1.4 & $4.3\times 10^{-5}$\\ %%4.260e-05 \\
\hline
\end{tabular}
\end{center}
Note: The values of $\kappa_{158}$ and $\beta_\mathrm{IR}$ are taken from \citet{Hirashita:2014a}.
\end{table}

As mentioned above, we exclude highly dust-obscured objects, which cannot be selected as LBGs. This means that IRX is not extremely large.
Thus, we impose a maximum IRX, IRX$_\mathrm{max}$, which is a free parameter in this paper.
In what follows, we explain how to calculate the IR and UV luminosities, and IRX.

The IR luminosity (denoted as $L_\mathrm{IR}$) is evaluated by
\begin{equation}
L_\mathrm{IR}\equiv\int_{0}^{\infty}L_{\nu}\, \mathrm{d}\nu
= C_\mathrm{IR}(M_\mathrm{d}/\mathrm{M_{\sun}})
(T_\mathrm{d}/\mathrm{K})^{\beta_\mathrm{IR}+4}~\mathrm{L}_{\sun},
%%=\int_{0}^{\infty}4\pi \kappa_{\nu(1+z)}M_dB_{\nu(1+z)}(T_{dust})\, d\nu
\end{equation}
where $C_\mathrm{IR}$ can be numerically evaluated by integrating
equation (\ref{eq:L_IR}) together with equation (\ref{eq:kappa}) for $\kappa_\nu$. The obtained value of $C_\mathrm{IR}$ is given for each dust species in Table \ref{tab:param}.

To calculate the UV luminosity (denoted as $L_\mathrm{UV}$), we assume that $L_\mathrm{UV}$ is converted from $M_\star$ with a factor $\alpha$:
\begin{align}
L_\mathrm{UV}=\alpha M_\star .\label{eq:LUV}
\end{align}
%%Here we assumed that the dust extinction plays a minor role in shaping the relation between $M_\star$ and $L_\mathrm{UV}$, which is consistent with our targets (LBGs) that are not highly obscured by dust.

We take the standard value of $\alpha$ (denoted as $\alpha_0$) from \citet{Schaerer:2015a}
\citep[see also][]{Wang:2017a}, who derived an almost linear relation between $L_\mathrm{UV}$ and $M_\star$ for LBGs at $z\sim 6.7$: $\alpha_0 =15.6 (\mathrm{L}_{\sun}/\mathrm{M}_{\sun})$ based on their value at $M_\star =10^9$ M$_{\sun}$. We confirm that this value of $\alpha_0$ is consistent with the relation between $L_\mathrm{UV}$ and $M_\star$ derived from SED fitting within a factor of $\sim 3$ for the major part of \cite{Burgarella:2020a}'s sample [except for a couple of low-$M_\star$ ($<10^9$ M$_{\sun}$) objects with $\alpha\gtrsim 100(\mathrm{L}_{\sun}/\mathrm{M}_{\sun})$ because of young ($\sim 10^7$ yr) stellar ages; we separately discuss high $\alpha$ in Section \ref{subsubsec:var_alpha_IRX}].
Considering the above factor 3 variation, we give $\alpha$ for each object as $\alpha =10^\delta\alpha_0$, where $\delta$ is randomly chosen from $[-0.5,\, 0.5]$. In reality, $\alpha$ depends on the dust extinction and the stellar age, but we avoid including this complication in our model in order to keep the simplicity. Thus, we take the above approach; that is, we choose $\alpha$ randomly for each object (by implicitly assuming that the physical parameters regulating $\delta$ vary randomly).

Using the above $L_\mathrm{IR}$ and $L_\mathrm{UV}$, we obtain IRX (noting that
$\mathcal{D}_\star =M_\mathrm{d}/M_\star$) as
\begin{align}
\mathrm{IRX}=C_\mathrm{IR}\mathcal{D}_\star T_\mathrm{d}^{\beta_\mathrm{IR}+4}/\alpha .\label{eq:IRX}
\end{align}
If $\mathrm{IRX}$ is larger than a certain threshold $\mathrm{IRX_{max}}$, we regard this object as a highly obscured galaxy,
and remove it from the sample.
We adopt $\mathrm{IRX}_\mathrm{max}\lesssim 10$ based on
actually observed values \citep{Bouwens:2016a,Burgarella:2020a} and further discuss it in Section \ref{subsec:IRXeffect}.

\subsection{Parameter setup} 
\label{subsec:parameter}
In our model, we give $(M_\star,\,\mathcal{D}_\star,\, T_\mathrm{d}^0)$ for each galaxy.
Since the statistical distribution of these quantities are poorly known,
we select the values of these quantities randomly in the ranges by referring to
actually observed or theoretically expected for high-redshift LBGs.
For the stellar mass and dust-to-stellar mass ratio, we
refer to \citet{Burgarella:2020a} and \citet{Nanni:2020a} for the ranges and adopt
$\log (M_\star /\mathrm{M}_{\sun})=8.0$--10.0 and $\log\mathcal{D}_\star =(-4)$--$(-1.5)$. \citet{Pozzi:2021a} derived $M_\mathrm{d}\lesssim 10^{7.5}$ M$_{\sun}$ for $M_\star\sim 10^{9.2}$--$10^{9.9}$ M$_{\sun}$ [converted from the observed UV luminosities using equation \ref{eq:LUV} with $\alpha =15.6(\mathrm{L}_{\sun}/\mathrm{M}_{\sun})$] at $z\sim 5$, indicating $\log\mathcal{D}_\star\lesssim (-1.7)$--$(-2.4)$. This is consistent with the adopted range of $\mathcal{D}_\star$.
For the stellar mass range, there are galaxies with $M_\star >10^{10}$ M$_{\sun}$ at high redshift, but they usually belong to populations different from LBGs, such as SMGs \citep[e.g.][]{Michalowski:2017a}.
We choose a logarithmic values for $M_\star$ and $\mathcal{D}_\star$ randomly from the above ranges.

For the dust temperature, since our purpose is to clarify the dust temperature bias,
we slightly extend the range from 40--70 K, which is derived by \citet{Burgarella:2020a}.
This temperature range covers the dust temperatures estimated with various methods
\citep{Faisst:2017a,Hashimoto:2019a,Inoue:2020a,Sommovigo:2021a,Bakx:2021a}.
We extend the range towards both lower and higher temperatures and adopt $T_\mathrm{d}^0=20$--85 K.
\citet{Burgarella:2020a} showed in their fig.\ 1 that some LBGs may be consistent with $T_\mathrm{d}=85$ K if we take the uncertainties into account.
\citet{Bakx:2020a} obtained a constraint for the dust temperature of an LBG at $z=8.3$ as $>80$ K, which justifies the above extension to a high dust temperature.
A SMG AzTEC-3 has a dust temperature of 92 K \citep{Riechers:2020a}: Although we exclude SMGs from our modelling, this object demonstrates a possibility that some high-redshift galaxies may have an extremely high dust temperature.
Numerical simulations, on the other hand, show moderate dust temperatures lower than 40 K
\citep{Ma:2019a,Liang:2019a}. Because we do not know the real dust temperature range, we extend it down to 20 K, which is a typical dust temperature in nearby star-forming galaxies \citep[e.g.][]{Draine:2007a}. The lowest dust temperature is not very important since the CMB limits the lowest temperatures achieved at high redshift (equation \ref{eq:CMB}).
However, we should note again that the wide temperature range is adopted for the purpose of examining possible temperature biases. We leave the determination of the correct temperature range for a future study because we need a larger observational sample with a uniform sensitivity and a further development of a dedicated statistical tool.

For the redshift, we examine $z=5$, 7, and 10. Including $z=10$ is useful to discuss the possibility of expanding the redshift frontier of the current observations.
The number of generated objects is adjusted to obtain statistically meaningful results as described below.

%%{\color{red}Considering CMB effect, the random temperatures adjusted by applying \cite{da-Chnha:2013a}method. Therefore, the result shown have already included the lower temperature limit for CMB effect at corresponding redshift.}
%%At $z>6.4$, the CMB temperature becomes higher than 20 K. In our model, we simply ignore the objects whose dust temperature is below the CMB temperature ($T_\mathrm{d}<2.7(1+z)$ K). 
%%We generate 3000 objects for the set of the three parameters.The number of objects is adjusted to obtain statistically meaningful results.

\subsection{Selection of LBGs detectable with ALMA}
Our main purpose is to examine the dust temperature bias for the detected objects.
For a representative sensitivity, we consider ALMA observations.
Although our quantitative conclusions are only valid for ALMA, the same biases are qualitatively expected for other (including future) submm telescopes. As mentioned in the Introduction, we use Band 6, 7 and 9 of ALMA in this study. Band 8 ($\sim 750~\micron$) is also used for the studies of high-redshift galaxies \citep[e.g.][]{Faisst:2020a,Inoue:2020a}, but Band 8 gives similar results to Band 7 for the diagrams we show below. Thus, we omit Band 8 for the conciseness of presentation.
With the ALMA sensitivity calculator,\footnote{\url{https://almascience.nao.ac.jp/proposing/sensitivity-calculator}}
the 3-$\sigma$ detection limits with 1 hour integration time at 450, 850, and 1200 $\micron$
(Band 9, 7, and 6, respectively) are 0.81, 0.088, and 0.063 mJy, %%0.808 mJy, 0.0877 mJy, and 0.0628 mJy
respectively. We also examine a deeper observation with 5 hour integration with 3-$\sigma$ limits of 0.36, 0.039, and 0.028 mJy %%0.3613 mJy, 0.0392 mJy, 0.0281 mJy
at 450, 850, and 1200 $\micron$, respectively. If we aim at 5-$\sigma$ detection with the same sensitivities, we require an integration time roughly 3 times longer.

\section{Result}\label{sec:result}
\subsection{Effect of the criterion on IRX}\label{subsec:IRXeffect}

\begin{figure}
\includegraphics[width=0.45\textwidth]{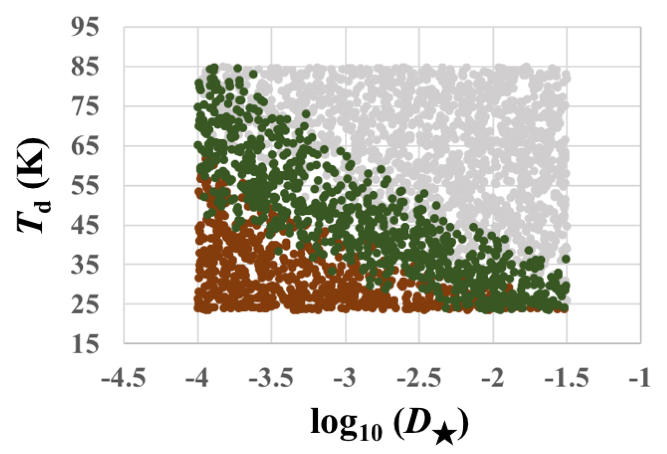}
\includegraphics[width=0.45\textwidth]{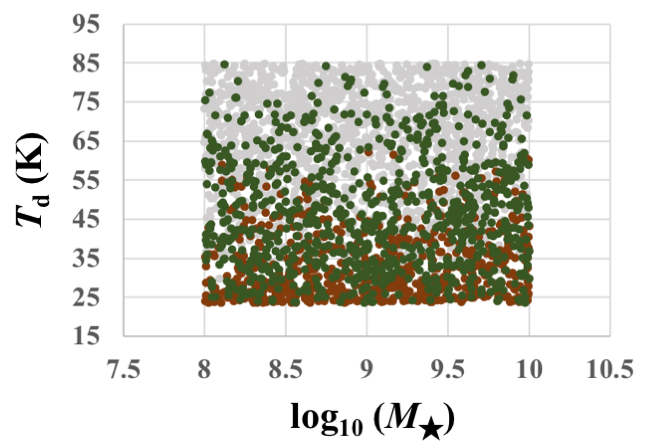}
\caption{Effects of the selection criterion $\mathrm{IRX}<\mathrm{IRX}_\mathrm{max}$
in the basic parameter space $(M_\star,\, \mathcal{D}_\star ,\, T_\mathrm{d})$.
The data points in the 3-dimensional space are projected onto the $T_\mathrm{d}$--$\mathcal{D}_\star$ (upper) and $T_\mathrm{d}$--$M_\star$ planes (lower). 
The data points in colour are selected with $\mathrm{IRX}_\mathrm{max}=10$, while those in brown are with $\mathrm{IRX}_\mathrm{max}=1$; that is, the green points indicate objects with $1\leq\mathrm{IRX}<10$. The grey points show the data with $\mathrm{IRX}\ge 10$; thus, they are not used in the analysis in this paper. }
\label{fig:IRX_selection}
\end{figure}

Before showing our results, we examine the effect of imposing the condition $\mathrm{IRX}<\mathrm{IRX}_\mathrm{max}$ (Section \ref{subsec:model_LIR}).
We generate 3,000 LBGs.
In Fig.\ \ref{fig:IRX_selection}, we show how the criterion for IRX affects the sample properties in the parameter space. The 3-dimensional parameter space is projected onto the $T_\mathrm{d}$--$\mathcal{D}_\star$ and $T_\mathrm{d}$--$M_\star$ planes.
We show the data at $z=7$; however, except for the lowest dust temperature determined by the CMB temperature, this figure does not depend on the redshift.

We observe that the $\mathrm{IRX}$ value mainly constraints the higher end of
the dust temperature and that the highest dust temperature decreases if we impose a severer upper limit on $\mathrm{IRX}$ (i.e.\ smaller IRX$_\mathrm{max}$). The upper bound of dust temperature strongly depends on $\mathcal{D}_\star$.
For a higher dust abundance ($\mathcal{D}_\star$), a high IR luminosity is more easily achieved, so that the dust temperature is strongly constrained from the upper limit of $\mathrm{IRX}$. From equation~(\ref{eq:IRX}), the boundary is described by $\mathcal{D}_\star T_\mathrm{d}^{\beta_\mathrm{IR}+4}=\alpha\mathrm{IRX}_\mathrm{max}/C_\mathrm{IR}$.
We regard the $\mathcal{D}_\star$-dependent upper bound for $T_\mathrm{d}$ as a physically reasonable
constraint, since, with a limited amount of dust-heating sources (stars), a large amount of
dust cannot be equally heated to a high temperature.
%%If a large amount of dust is present, (i) the dust temperature is low because of the limited amount of heating or (ii) only a part of dust is heated. The case (i) is shown by the low-$T_\mathrm{d}$ limit for high $\mathcal{D}_\star$ in Fig.\ \ref{fig:IRX_selection} (upper), while the case (ii) is represented effectively by a case of low $\mathcal{D}_\star$.
In other words, $\mathcal{D}_\star$ and $T_\mathrm{d}$ are not completely independent. Therefore, we do not use $\mathcal{D}_\star$ explicitly as an independent variable in statistical discussions in Section \ref{sec:discussion}.
In the $T_\mathrm{d}$--$M_\star$ diagram, there is no clear temperatures trend along the $M_\star$ axis. The objects occupy both high and low dust temperature regions at any $M_\star$.

We hereafter adopt $\mathrm{IRX}_\mathrm{max}=10$. Thus, galaxies with $\mathrm{IRX}>10$ shown in grey in Fig.\ \ref{fig:IRX_selection} are removed from the sample. In the end, we have 1,510 objects with $\mathrm{IRX}\leq 10$.
If we adopt $\mathrm{IRX}_\mathrm{max}=1$, the number of high-$T_\mathrm{d}$ objects decreases. 
There are some observational clues to the maximum value of IRX at $z>5$.
\cite{Fudamoto:2020a} using stacked data showed that the IRX is typically smaller than 1 at $z=5.5$
in the stellar mass range we consider.
However, the stacked data do not constrain the maximum $\mathrm{IRX}$.
They also showed detected data points around $M_\star\sim10^{10}M_{\sun}$ with $\log\mathrm{IRX}\sim 0.5$, implying the existence of objects with $\mathrm{IRX}>1$.
\citet{Hashimoto:2019a}\footnote{Some papers define $\log\mathrm{IRX}$ as IRX.} estimated that some LBGs at $z\gtrsim 7$ have $\mathrm{IRX}=1$--10.
Since ALMA-detected LBGs are rare at $z>5$, simulations also help to derive expected ranges of IRX:
\citet{Ma:2019a} showed that IRX extends up to 10 at $M_\star\gtrsim 10^9$ M$_{\sun}$ \citep[see also][]{Vijayan:2021a}. However, $\mathrm{IRX}\lesssim 1$ at $M_\star <10^9$ M$_{\sun}$.
Thus, applying $\mathrm{IRX}_\mathrm{max}=10$ could overestimate the detectability of objects with
$M_\star$ less than $10^9$ M$_{\sun}$. However, we show below that such low-$M_\star$ LBGs are hardly detected even with $\mathrm{IRX}_\mathrm{max}=10$. Thus, the possibility of lower IRX for low-$M_\star$ LBGs
does not affect our conclusions.
We still mention the results with $\mathrm{IRX}_\mathrm{max}=1$ in Section \ref{subsubsec:var_alpha_IRX},
but we focus on the calculations of $\mathrm{IRX}_\mathrm{max}=10$ unless otherwise stated.

\subsection{Characteristics of detected objects}
Now we examine the detectability by ALMA for the selected sample. We particularly examine in which sense the detected objects are biased in terms of the dust temperature.

\subsubsection{$T_\mathrm{d}$ vs.\ $L_\mathrm{IR}$}\label{subsubsec:L_IR}
\begin{figure*}
\includegraphics[width=0.9\textwidth]{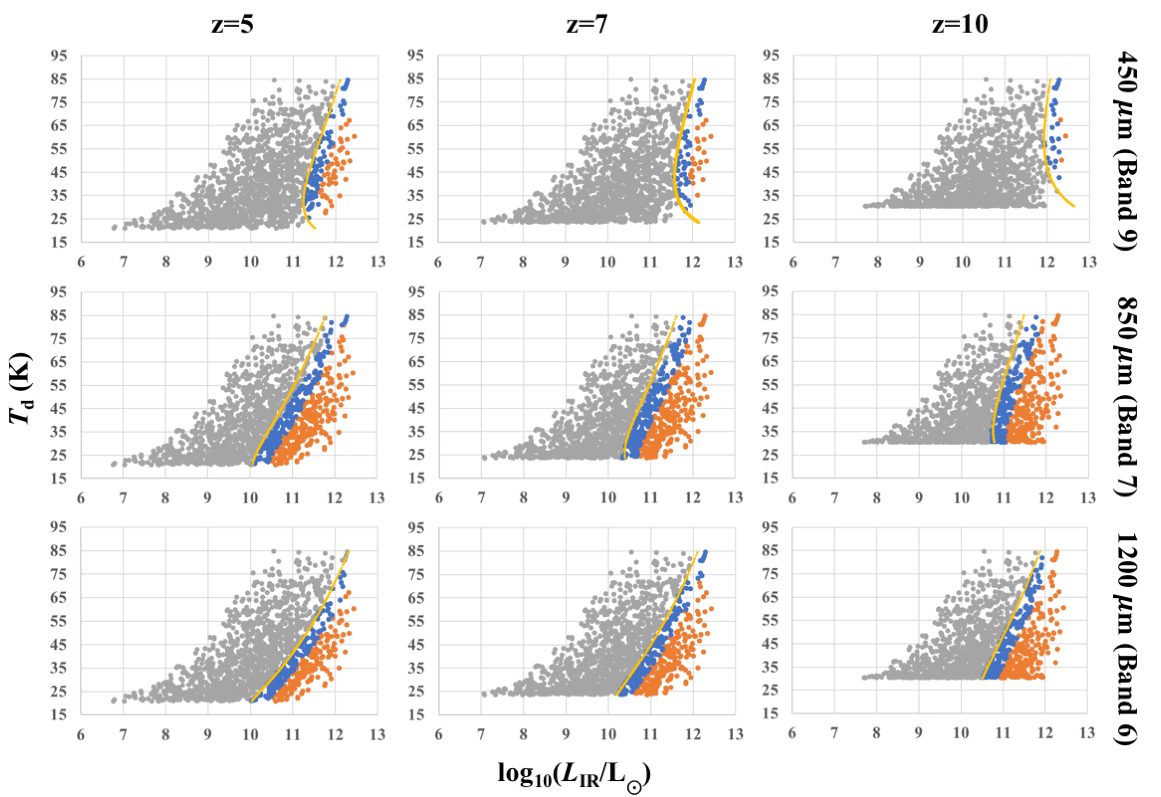}
\caption{$T_\mathrm{d}$ vs.\ $L_\mathrm{IR}$ for the three different redshifts, $z=5$, 7, and 10 at 450, 850, and 1200 $\micron$ (ALMA Band 9, 7, and 6, respectively) as indicated on the top and on the right side. Only objects with $\mathrm{IRX}\leq 10$ are plotted. The orange and blue points represent galaxies which can be detected (3$\sigma$) with integration time $<1$ hour and 1--5 hours with ALMA, respectively. The grey points indicate objects not detected with 5 hours of integration. The yellow solid line is the analytically calculated detection limit corresponding to the ALMA 5-hour integration (3$\sigma$).}
\label{fig:LIR_2.0}
\end{figure*}

First we focus on the two quantities related to dust emission: $T_\mathrm{d}$ and $L_\mathrm{IR}$.
In Fig.~\ref{fig:LIR_2.0}, we show the distribution of the sample on the $T_\mathrm{d}$--$L_\mathrm{IR}$ diagram.
Since the detectability depends on the observational band and the redshift,
we separately plot nine panels for $\lambda =450$, 850, and 1200~$\micron$
(ALMA Band 9, 7, and 6, respectively) and $z=5$, 7, and 10.
Note that the distribution of the data points is almost the same for all the panels,
and the only difference among various redshifts is caused by the lower bound of the dust temperature constrained by the CMB temperature.
Since the SED is determined by $T_\mathrm{d}$ and $L_\mathrm{IR}$, it is possible to calculate the detection limit analytically as shown by the solid yellow curves.
As expected, the detected objects have larger $L_\mathrm{IR}$.
The number of detectable objects at 450~$\micron$ is significantly smaller than those at the other wavelengths. Moreover, at $z=10$,
%%說明z10 在450很少 且在850 1200 比z7 少
the wavelength where the SED peaks shifts beyond 450~$\micron$ (i.e.\ 450~$\micron$ is located on Wien's side) so that the detection becomes significantly difficult. In contrast, the detection at 850 and 1200 $\micron$ is not sensitive to the redshift because of the so-called negative $K$ correction. The difference in the number of detected objects at 850 and 1200 $\micron$ is only 10 per cent. All the objects detectable at 450 $\micron$ can be detected at 850 and 1200 $\micron$.

We also observe that the detection is sensitive to the dust temperature
at 850 and 1200~$\micron$ while it is less so at 450~$\micron$.
With the same $L_\mathrm{IR}$, lower-$T_\mathrm{d}$ objects are more easily detected at 850 and 1200 $\micron$: Objects with $L_\mathrm{IR}\sim\mbox{a few}\times 10^{10}$ L$_{\sun}$ can be detected if $T_\mathrm{d}\lesssim 30$ K, and only very IR luminous objects with $L_\mathrm{IR}\gtrsim 10^{11}$ L$_{\sun}$ can be detected if $T_\mathrm{d}\gtrsim 45$ K. With a fixed total IR luminosity, the SED peak shifts towards a longer wavelength for lower $T_\mathrm{d}$, so that the detection at wavelengths 850 and 1200 $\micron$, which are mostly located on the Rayleigh--Jeans side, becomes easier. This is the reason why the detected objects at 850 and 1200 $\micron$ are biased towards low dust temperature \citep[e.g.][]{Chapman:2005a}, as further discussed in Section \ref{subsec:prob}.
In contrast, the boundary of the detected objects (yellow solid line) is less inclined in the 450 $\micron$ band, since it is located around the SED peak at $z\sim 5$--7. Thus, the change of dust temperature has a smaller influence on the detectability in the 450 $\micron$ band than in the longer-wavelength bands. As shown later, the 450 $\micron$ band is not bias-free, but the 450 $\micron$ sample covers the entire range of dust temperature. Moreover, all objects detected at 450 $\micron$ are detected at 850 and 1200 $\micron$.
%%Indeed, at 450 $\micron$, almost all objects regardless of the dust temperature are detected as long as $\log (L_\mathrm{IR}/\mathrm{L}_{\sun})\gtrsim 11.5$ and 11.8 for 5- and 1-h integration, respectively.
Therefore, a survey at such a short wavelength as 450 $\micron$ is useful to construct a (sub)sample, whose dust temperature is known. In this sense, a survey data at 450 $\micron$ would be useful to overcome the dust temperature bias at $z\sim 5$--7. The sample size is yet limited by the shallowness of the 450 $\micron$ observations. This means that a deeper survey at 450 $\micron$ is crucial to further increase the size of the sample whose dust temperature is known. We discuss a possibility of a deeper survey in Section \ref{subsec:450}.

%% Td vs Ds
%%\textbf{Probably, this section and Fig. 3 are not necessary because we say in the above that we don't use
%%$\mathcal{D}_\star$ explicitly. Instead, in the next section, we also show the $T_\mathrm{d}$--$M_\star$ diagram (at $z=7$) for 450 and 850 $\micron$ only for the detected objects colour-coded with $\mathcal{D}_\star$, in order to show that the detection is driven by the dust abundance.}

%%Next we show the distribution of the sample data on the $T_\mathrm{d}$--$\mathcal{D}_\star$ diagram in Fig.\ \ref{fig:D_star}.
%%The distribution of the points is the same as in Fig.\ \ref{fig:IRX_selection}, but we removed the data with $\mathrm{IRX}>10$.
%%The upper bound of $T_\mathrm{d}$ systematically decreases as $\mathcal{D}_\star$ decreases, simply because we impose the upper limit for IRX as shown in Fig.\ \ref{fig:IRX_selection} (i.e.\ dust-rich high temperature objects have too high values of IRX).
%%As $\mathcal{D}_\star$ decreases, only objects with high $T_\mathrm{d}$ can be detected by ALMA, because high-$T_\mathrm{d}$ objects emit more efficiently. However, the detectability is not simply determined by the dust temperature: some high-$T_\mathrm{d}$ objects are not detected by ALMA, since the total dust mass is determined not only by $\mathcal{D}_\star$ but also by $M_\star$. Thus, if $M_\star$ is small, the submm fluxes become smaller (see also Section \ref{subsubsec:M_star}).
%%The difference in the detectability among the redshifts and bands is the same as discussed above in Section \ref{subsubsec:L_IR}.

\subsubsection{$T_\mathrm{d}$ vs.\ $M_\star$ ($L_\mathrm{UV}$)}\label{subsubsec:M_star}

\begin{figure*}
\includegraphics[width=0.9\textwidth]{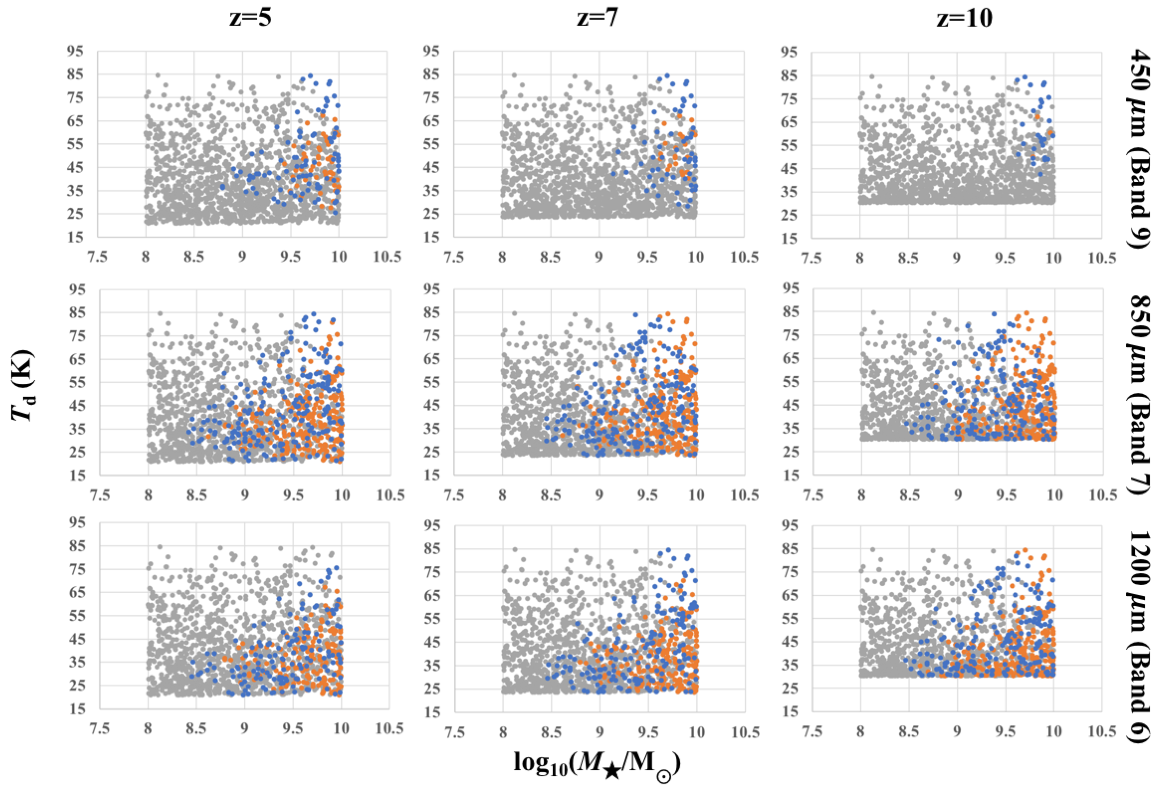}
\caption{Same as Fig.\ \ref{fig:LIR_2.0} but showing the $T_\mathrm{d}$--$M_\star$ diagrams.}\label{fig:Ms_2.0}
\end{figure*}

Next we show the $T_\mathrm{d}$--$M_\star$ relations in Fig.~\ref{fig:Ms_2.0}. We observe that there is a tendency that objects with larger $M_\star$ are more easily detected. However, a high stellar mass does not necessarily lead to detection. This is because of the difference in the dust abundance ($\mathcal{D}_\star$) (see Section \ref{subsec:IRXeffect}).
%%說明多數集中在高Ms, 但高Ms 不是全部都可以被偵測，因為還是一個隱藏參數
As mentioned above, the detection at 450~$\micron$ is extremely hard at $z=10$. For $z=5$ and 7,
detected galaxies at 450 $\micron$ have $\log (M_\star/\mathrm{M}_{\sun})\gtrsim 9.5$, while objects with a wider variety in $M_\star$ ($\gtrsim 10^{8.5}$ M$_{\sun}$) are detected at 850 and 1200 $\micron$. This difference reflects the different depths.
At 450 $\micron$, although there is a tendency that galaxies with high $M_\star$ are more easily detected, there is not a clear trend that higher- or lower-$T_\mathrm{d}$ objects are more easily detected for a fixed $M_\star$ at $T_\mathrm{d}\gtrsim 30$ K.
At 850 and 1200 $\micron$, lower-$T_\mathrm{d}$ objects ($T_\mathrm{d}\lesssim 40$ K) are more easily detected at low $M_\mathrm{\star}\sim 10^9~\mathrm{M}_{\sun}$ (further discussed in Section \ref{subsec:prob}). 
For the difference in the detectability among the redshifts and bands, see Section \ref{subsubsec:L_IR}.

\subsection{Detection probability}\label{subsec:prob}

In the above, we adopted two reference quantities: $L_\mathrm{IR}$ and $M_\star$. The detectability for various $T_\mathrm{d}$ is different depending on which of these two quantities is used for the sample selection.
We show the detectability as a function of $T_\mathrm{d}$ using one of the above two quantities ($L_\mathrm{IR}$ and $M_\star$) as a reference.
If we use $L_\mathrm{IR}$ and $M_\star$ for the reference quantity, we, respectively, refer to the sample as \textit{the IR-referenced sample} and \textit{the UV-referenced sample} (recalling that $M_\star$ is originally derived from the UV luminosity by assumption).

Here we show the detection probability, which is defined as the fraction of
the detectable objects to the generated sample in each of the bins set below.
The above sample size is still small to show the statistical properties for various bins of dust temperature, stellar mass, and IR luminosity. Thus, for the purpose of showing the detection probability, we boost the sample by ten times with the same procedure as described in Section \ref{subsec:parameter}.
We adopt the 5-hour detection limit (3 $\sigma$) for ALMA.
%%The reason for scaling up the sample number up to 30000 is in order to obtain more precise probabilities by shrinking error bars.
The objects are divided into 6 bins for the dust temperature with 10 K difference from 25 to 85 K.
%%(25--35, 35--45, 45--55, 55-65, 65-75 and 75-85K).
We also divide $\log L_\mathrm{IR}$ into 3 bins with a bin size of 0.5 dex in $\log(L_\mathrm{IR}/\mathrm{L}_{\sun})\ge 11$, and $M_\star$ into 4 bins with a bin size of 0.5 dex in $\log(M_\star/\mathrm{M}_{\sun})=8$--10.
Since no objects can be detected at 450 $\micron$ at $z=10$, we only show the results %%for 450, 850 and 1200 $\micron$
at $z=5$ and 7. 

\subsubsection{IR-referenced sample}
\label{subsubsec:LIR-selected sample}

\begin{figure*}
\includegraphics[width=0.8\textwidth]{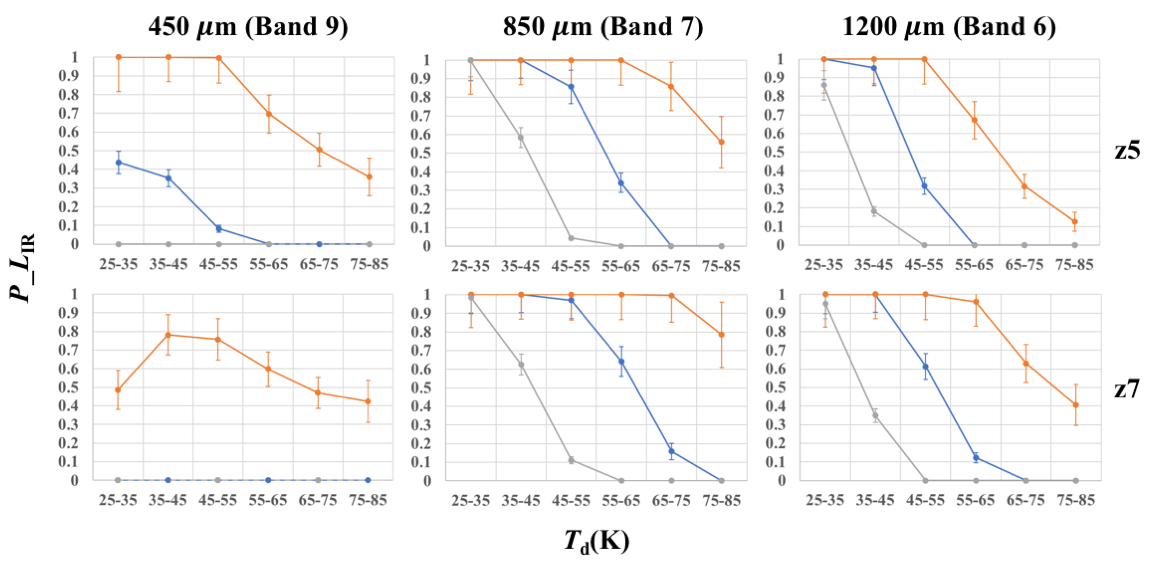}
\caption{Detection probability as a function of dust temperature (binned with 10-K width) for various ranges of $L_\mathrm{IR}$. The left, middle, and right panels show the results in the 450, 850, and 1200 $\micron$ bands
(ALMA Band 9, 7, and 6), respectively, and the upper and lower panels present $z=5$ and 7, respectively. The grey, blue and orange lines represents the sample with conditions $10.5<\log(L_\mathrm{IR}/\mathrm{L}_{\sun})<11$, $11<\log(L_\mathrm{IR}/\mathrm{L}_{\sun})<11.5$ and $11.5<\log(L_\mathrm{IR}/\mathrm{L}_{\sun})$, respectively.
%%The dash line represents two lines combination.
Note that objects with $\log (L_\mathrm{IR}/\mathrm{L}_{\sun})<11$ ($<11.5$) are not detected
in the 450 $\micron$ band at $z=5$ (7).}
The bars show the Poisson errors.
\label{fig:P_LIR}
\end{figure*}

We show the resulting detection probability for various ranges of $L_\mathrm{IR}$
in Fig.~\ref{fig:P_LIR}.
The detection probabilities are shown separately for $\log( L_\mathrm{IR}/\mathrm{L}_{\sun})=10.5$--11, 11--11.5 and $>$11.5; note that we only show $\log( L_\mathrm{IR}/\mathrm{L}_{\sun})>11$ for 450 $\micron$ at $z=5$ and $\log (L_\mathrm{IR}/\mathrm{L}_{\sun})>11.5$ for 450 $\micron$ at $z=7$ because less luminous galaxies are not detected.
If the IR luminosity is high, the detection probability is high as expected. However, even for $\log(L_\mathrm{IR}/\mathrm{L}_{\sun)}>11$, the detection is strongly biased towards lower $T_\mathrm{d}$ at all wavelengths at both $z=5$ and 7. In particular, the bias is sharp at 850 and 1200~$\micron$ in the sense that almost all objects are detected at low $T_\mathrm{d}$ while almost none is detected at high $T_\mathrm{d}$. The dust temperature at which this transition from detection to non-detection occurs depends on $L_\mathrm{IR}$ with higher $L_\mathrm{IR}$ allowing detection up to higher dust temperature. At 450~$\micron$, the decline of the detection probability towards higher $T_\mathrm{d}$ is milder, although the sensitivity is less than the other bands. This is consistent with the discussion in Section \ref{subsubsec:L_IR}. Although the 450 $\micron$ sample is useful to determine the dust temperature, the sample size is made small by requiring the detection at 450 $\micron$. Deeper 450 $\micron$ surveys in the future will be useful as we discuss further in Section \ref{subsec:450}.
%%{\color{red}This means that  the sample constructed based on the 450 $\micron$ detection is useful to avoid the dust temperature bias and that the bias should be somehow corrected for 850 $\micron$ (millimetre)-selected sample.}

\subsubsection{UV-referenced sample}\label{subsubsec:Ms-selected}

\begin{figure*}
\includegraphics[width=0.8\textwidth]{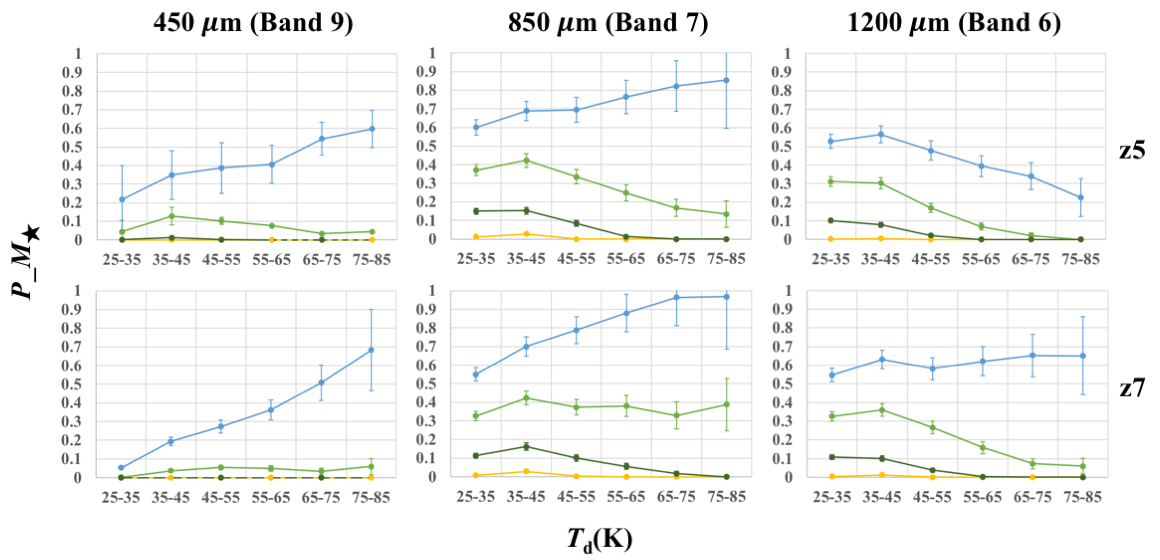}
\caption{Same as Fig.\ \ref{fig:P_LIR} but based on the UV ($M_\star$) selection. The yellow, dark green, light green and blue lines represent the sample with conditions $8<\log (M_\star/\mathrm{M}_{\sun})<8.5$, $8.5<\log (M_\star/\mathrm{M}_{\sun})<9$, $9<\log(M_\star/\mathrm{M}_{\sun})<9.5$ and $9.5<\log(M_\star/\mathrm{M}_{\sun})<10$, respectively. Note that objects with $\log (M_\star/\mathrm{M}_{\sun})<9$ are not detected in the 450 $\micron$ band at $z=7$.}\label{fig:P_M}
\end{figure*}

We show the detection probability for various ranges of $M_\star$ in Fig.~\ref{fig:P_M}.
The detection probabilities are shown separately for
$\log( M_\star /\mathrm{M}_{\sun})=8$--8.5, 8.5--9, 9--9.5, and 9.5--10.
Note that at 450 $\mu$m, only objects with $\log (M_\star /\mathrm{M}_{\sun}) >9$ (9.5)
are detected at $z=5$ (7).
From Fig.~\ref{fig:P_M}, we observe that the temperature bias depends on the stellar mass. If the stellar mass is high, there is no significant dust temperature bias at 450 $\micron$ at $z=5$ and 1200 $\micron$ at $z=7$.
There is a slight trend that higher-$T_\mathrm{d}$ objects are more easily detected in the 450~$\micron$ band at $z=7$ while lower-$T_\mathrm{d}$ objects are more detectable at 1200 $\micron$. At 850 $\micron$, LBGs with $\log (M_\star /\mathrm{M}_{\sun}) >9.5$ are mostly detected with a slight bias towards high-$T_\mathrm{d}$ objects. Less UV-luminous objects with $\log (M_\star /\mathrm{M}_{\sun})<9.5$ are biased towards low $T_\mathrm{d}$ for the 850 $\micron$ band in a similar way as observed for 1200 $\micron$. These complex behaviours of the bias is due to the two completing effects: if the dust temperature is high, only low-dust-abundance objects are permitted as described in Section \ref{subsec:IRXeffect} (Fig.\ \ref{fig:IRX_selection}). Thus, more efficient dust emission with higher $T_\mathrm{d}$ competes with less dust with lower $\mathcal{D}_\star$.
%%As a corollary, the detection rate is regulated by the dust abundance ($\mathcal{D}_\star$).

The low detection rate of dust emission from LBGs at $z>5$ by ALMA observations
\citep[e.g.][]{Capak2015a,Bouwens:2016a} is consistent with the low (half or less) detection probabilities in Band 6 (1200 $\micron$; Fig.\ \ref{fig:P_M} right), which is often used to observe dust emission at high redshift. More quantitatively, in a recent sample from \citet{Schouws:2021a}, who targeted LBGs at $z\gtrsim 7$ mostly with $\log M_\star /\mathrm{M}_{\sun}\gtrsim 9.5$,
6 out of 15 LBGs are detected at $\sim 1200~\micron$. The detection rate is $\sim 40$ per cent.
Their integration time per object is roughly 1 h; if we use the detection limit for 1 h in our model, the detection probability becomes roughly half. Since our model predicts a detection rate of $\sim 60$ per cent at $z=7$ in the 1200 $\micron$ band (Fig.\ \ref{fig:P_M}), the above detection rate (40 per cent) is reasonable considering their shallower detection limit.
\citet{Fudamoto:2020a} showed the detection rates in Band 7 as a function of stellar mass for galaxies at $z\sim 5$. Their on-source integration time is on average $\sim$1/3 h. Thus, their detection limit is roughly 4 times higher than that used above.
In this case, the detected fraction is $\sim 5$ times smaller. Therefore, in the highest mass range ($M_\star =10^{9.5}$--$10^{10}$ M$_{\sun}$), we predict a detection rate of $\sim 10$ per cent, which is consistent with their detection rate. In the range of $M_\star =10^9$--$10^{9.5}$ M$_{\sun}$, the detection rate is predicted to be a few per cent, which is also consistent with their extremely low detection rate.
Although our model needs further refinement for detailed comparison with observations, this broad success in explaining the detection rates supports our modelling in this paper.

%%Given that the IR-selected sample above has a clear $T_\mathrm{d}$ bias, it is safer to discuss the evolution of dust temperature using a UV-selected sample. Fortunately, LBGs are by construction selected by UV luminosity, especially at $z\gtrsim 5$.
For the current sample of $z>5$ LBGs detected by ALMA, the stellar masses are broadly larger than $10^{9.5}$ M$_{\sun}$ (or the UV luminosity higher than $10^{10.7}$ L$_{\sun}$ from equation \ref{eq:LUV}) and the detection is mostly based on Band 6 ($\lambda\sim 1200~\micron$) or multiple bands including Band 6 \citep{Willott:2015a,Faisst:2020a,Schouws:2021a}.
According to Fig.\ \ref{fig:P_M}, the detection at $z=5$--7 is not significantly biased or is slightly biased towards low $T_\mathrm{d}$ in Band 6. Thus, the high dust temperatures (40--70 K; \citealt{Burgarella:2020a}; see also the Introduction) obtained from the observations cannot be due to a bias but are reflecting the real trend.

\subsubsection{Possible variations caused by $\alpha$ and $\mathrm{IRX}_\mathrm{max}$}\label{subsubsec:var_alpha_IRX}

Here we discuss how much the variations of $\alpha$ and $\mathrm{IRX}_\mathrm{max}$, which could have a large diversity or uncertainty at high redshift (Section \ref{subsec:model_LIR}), affect the above detection probabilities.

As discussed in Section \ref{subsec:model_LIR}, a minor fraction of %%low-$M_\star$ ($\lesssim 10^{9}M_{\star}$) 
galaxies have large values of $\alpha$ ($\gtrsim 100~ \mathrm{L}_{\sun} /\mathrm{M}_{\sun}$), which are out of the range we adopted. These very high values of $\alpha$ is predominantly due to extremely young ages ($\sim 10^{7}$ yr). To examine the effect of large $\alpha$ on the detection probabilities, we examine a case where $\alpha =100~\mathrm{L}_{\sun} /\mathrm{M}_{\sun}$ for all the generated sample. Since larger values of $\alpha$ mean larger $L_\mathrm{UV}$, objects with larger $L_\mathrm{IR}$ are permitted under a fixed value of $\mathrm{IRX}_\mathrm{max}$. Thus, for the IR-referenced sample, the number of high-$L_\mathrm{IR}$ objects increases, but the detection probability at a fixed $L_\mathrm{IR}$ bin is not sensitive to the increase of $\alpha$. In contrast, $\alpha$ directly affects the detection probabilities of the UV-referenced sample. The detection probabilities increase in any of the $M_\star$ bins broadly by a factor of $\sim 2$, but the trends among different dust temperatures and stellar masses are kept similar. Even with this extreme value of $\alpha$, the detection rate is less than 30 per cent for $M_\star<10^9$ M$_{\sun}$, for which large $\alpha$ is actually observed (Section \ref{subsec:model_LIR}). Considering that the large $\alpha$ only moderately affects the results, we argue that the discussions and conclusions in this paper are not much altered by the existence of objects with extremely large $\alpha$.

%%(written above)As we mention in the Section \ref{subsec: IRX effect}, high-$D_\mathrm{_\star}$ objects do not have enough dust-heating source, so dust temperatures trend is biased to low dust temperatures. High-$D_\mathrm{\star}$ objects which have low dust temperatures are easier to detect at low $M_\mathrm{\star}$ since we give the $M_\mathrm{d}$ value. Therefore, we can easily see the temperature trend is reversed between low and high $M_\mathrm{_\star}$.
%%As shown in Section \ref{subsubsec:M_star}, the detection probabilities are predominantly driven by the dust abundance (dust-to-stellar mass ratio). Thus, the absolute value of the detection probability depends on the actual distribution function of dust-to-stellar mass ratio.

For IRX, many LBGs may have lower values ($\sim 1$) as mentioned in Section \ref{subsec:IRXeffect}. Thus, we also examine a case of
$\mathrm{IRX}_\mathrm{max}=1$ instead of 10. We find that this case predicts completely suppressed detection of objects with $M_\star<10^{9}M_{\sun}$, which are difficult to detect even with $\mathrm{IRX}_\mathrm{max}=10$. The detection probability for $M_\star>10^{9.5}M_{\sun}$ drops by half. For the IR-referenced sample, high-$T_\mathrm{d}$ ($T_\mathrm{d}\gtrsim 60$K) and high-$L_\mathrm{IR}$ ($L_\mathrm{IR}\gtrsim 10^{11.5}L_{\sun}$) objects are eliminated because of the severer criterion for IRX.
%%since some LBGs around $M_\star\sim 10^{10}$ M$_{\sun}$ have $\mathrm{IRX}>1$ (see the discussion in Section \ref{subsec:IRXeffect}),
However, we regard $\mathrm{IRX}_\mathrm{max}=1$ an extreme assumption since some LBGs likely have $\mathrm{IRX}>1$
(Section \ref{subsec:IRXeffect}).

\section{Discussion}\label{sec:discussion}

\subsection{Different grain species}\label{subsec:species}
Although we only showed the results for graphite, we also analyzed different types of grains (AC and silicate; Table \ref{tab:param}). The temperature biases in the IR-referenced and UV-referenced samples are similar to what we have shown using graphite.
%%The detected galaxies number of silicate is slightly different to graphite within {\color{red}0.66$\%$ difference}.
The detected number is $\sim$ 26 (20) per cent larger for AC (silicate) than for graphite.
%%{\color{red}33$\%$ }difference to the number of graphite.
Both AC and silicate slightly extend the sample towards higher dust temperatures (by at most 10~K) because the constraint from $\mathrm{IRX}<\mathrm{IRX}_\mathrm{max}$ depends on $C_\mathrm{IR}$ and $\beta_\mathrm{IR}$ (equation \ref{eq:IRX}). Moreover, a smaller $\beta_\mathrm{IR}$ for AC enhances the emission at long wavelengths, leading to more detected objects at 850 and 1200 $\micron$. However, $C_\mathrm{IR}$ and $\beta_\mathrm{IR}$ do not modify the trend of the detection probability for $T_\mathrm{d}$, so that the above conclusions are not qualitatively affected. We just note that the detection probability could be affected at most by a factor of 1.3 if we adopt other dust species than graphite.

\subsection{Deeper 450 $\micron$ observations}
\label{subsec:450}
In the above, the detected objects show different biases for $T_\mathrm{d}$ between $450~\micron$ and 850 (or 1200) $\micron$.
However, the overall sample size is limited by the shallowness of the 450~$\micron$ band.
The 450~$\micron$ band, which is near the SED peak or even on Wien's side of the SED, is also crucial to determine the dust temperature.
A future large single-dish telescope in a site with a very low water vapour condition is expected to significantly improve the sensitivity at 450~$\micron$ (and in some shorter wavelength bands; nearly THz frequencies) compared with ALMA.
The Antarctic Plateau provides the most suitable atmospheric condition for
ground-based THz observations. Therefore, among various plans in the world,
we focus on a 30-m THz telescope at New Dome Fuji (PI: Naomasa Nakai).
We calculated the sensitivity using the following assumptions.
We adopt the transmission achieved in 50 per cent of the time in winter; that is,
0.75 at 650 GHz (450 $\micron$) obtained at Dome A.
%%(transmission of 0.90, 0.71, and 0.31 at 400, 850, and 1300 GHz, respectively,
The transmission at New Dome Fuji is expected to be almost the same as that at Dome A \citep{Yang:2010a}.
The precision of the antenna surface is 20 $\micron$.
For the sensitivity of the camera, we adopt $\mathrm{NEP}=6\times 10^{-18}$ W Hz$^{-1/2}$.
We obtain the expected detection limits ($5\sigma$) for a
10-hour on-source integration at 450 $\micron$ as 0.069 mJy.
One of the most important advantage of the 450 $\micron$ band compared with longer wavelengths is that the sensitivity is not confusion-limited because of a higher angular resolution. Therefore, large improvement of sensitivity at 450 $\micron$ is expected by future large single-dish telescopes if they are located in a site of good atmospheric condition.

We examine the same sample as in Section \ref{subsec:prob} but adopt the above deeper detection limit at 450 $\micron$. The results for the IR- and UV-referenced samples are described in what follows.

\subsubsection{IR-referenced sample}
\begin{figure}
\includegraphics[width=0.49\textwidth]{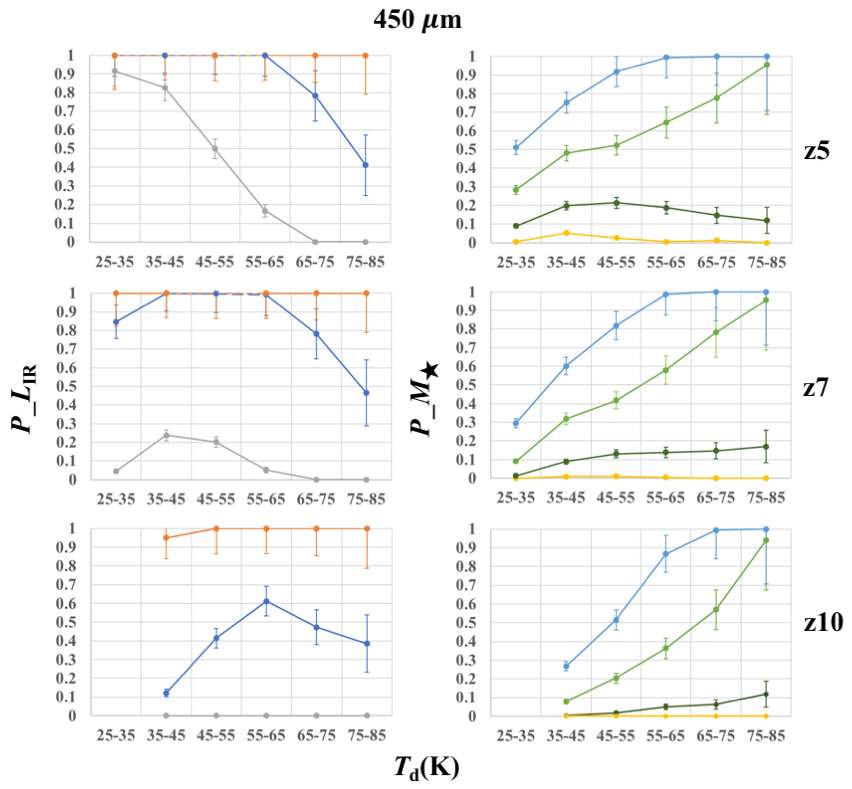}
\caption{Detection probabilities for future sensitive 450 $\micron$ observations described in the text. The left and the right columns of panels are the same as Figs.\ \ref{fig:P_LIR} and \ref{fig:P_M}, respectively, but only for 450 $\micron$ with the better sensitivity. Left: The grey, blue and orange lines represents the sample with conditions $10.5<\log(L_\mathrm{IR}/\mathrm{L}_{\sun})<11$, $11<\log(L_\mathrm{IR}/\mathrm{L}_{\sun})<11.5$ and $11.5<\log(L_\mathrm{IR}/\mathrm{L}_{\sun})$, respectively. Right: The yellow, dark green, light green and blue lines represent the sample with conditions $8<\log (M_\star/\mathrm{M}_{\sun})<8.5$, $8.5<\log (M_\star/\mathrm{M}_{\sun})<9$, $9<\log(M_\star/\mathrm{M}_{\sun})<9.5$ and $9.5<\log(M_\star/\mathrm{M}_{\sun})<10$, respectively. The upper, middle, and lower rows of panels show the results at $z=5$, 7, and 10, respectively. At $z=10$, we do not show the data in the temperature bin 25--35K because the CMB temperature lies in this range. Note that no LBG with $\log(L_\mathrm{IR}/\mathrm{L}_{\sun})<11$ is detected at $z=10$ and none with  $\log( M_\star/\mathrm{M}_{\sun})<8.5$ and $<9$ is detected at $z=7$ and 10, respectively.}\label{fig:450}
\end{figure}

In Fig.~\ref{fig:450} (left), we show the detection probability as a function of dust temperature for various ranges of $L_\mathrm{IR}$. Compared with Fig.~\ref{fig:P_LIR} (left), the detection probabilities rise significantly, especially for objects with $\log(L_\mathrm{IR}/\mathrm{L}_{\sun})>11$. Almost all LBGs with $\log(L_\mathrm{IR}/\mathrm{L}_{\sun})>11.5$ are detected at both $z=5$ and 7, and even objects with $11<\log(L_\mathrm{IR}/\mathrm{L}_{\sun})<11.5$ have twice higher detection probabilities than those by ALMA. Besides, some low-luminosity objects with $\log(L_\mathrm{IR}/\mathrm{L}_{\sun})<11$ can be detected. In summary, a nearly complete, IR-selected, sample can be constructed at $L_\mathrm{IR}>10^{11}$ L$_{\sun}$ for high-redshift ($z=5$--7) LBGs. As shown in Section \ref{subsubsec:L_IR}, detection becomes difficult for $z=10$ at 450 $\micron$ by ALMA.
From Fig.\ \ref{fig:450}, we observe for $z=10$ that the detection probability for objects with $\log(L_\mathrm{IR}/\mathrm{L}_{\sun})>11.5$ is higher than 0.8 and that with $11<\log(L_\mathrm{IR}/\mathrm{L}_{\sun})<11.5$ reaches almost half. Thus, we expect that more information for dust emission from LBGs at $z=10$ can be obtained with the future telescope.
%%Note that, since the 850 $\micron$ and 1200 $\micron$ bands are biased towards low $T_\mathrm{d}$, it is difficult to construct an IR-selected sample as is clear from Fig.\ \ref{fig:P_LIR}. Moreover,
At 850 and 1200 $\micron$, ALMA is still preferable since
a single-dish telescope survey becomes confusion-limited at such long wavelengths. Therefore, future 450 $\micron$ surveys with the 30-m-class single-dish Antarctic telescope, combined with ALMA measurements at longer submm wavelengths, will be promising not only to measure the dust temperature but also to construct a deeper IR-selected sample with little dust-temperature bias.

We also performed a calculation for 350 $\micron$ observations with the same future telescope but with a transmission of 0.71 (50 per cent in winter), and obtained detection probabilities similar to, but slightly worse than, the above 450 $\micron$ case. Thus, the 450 $\micron$ band is optimum for constructing an IR-selected LBG sample at $z=5$--7. Note that optical observations will also be advanced in the future, so that the cross-identification with optical telescopes will not be a limiting factor to identify LBGs.

\subsubsection{UV-referenced sample}
As shown in Fig.\ \ref{fig:450} (right), objects
with $\log (M_\star /\mathrm{M}_{\sun}) >9$ are biased towards high dust temperatures.
This is interpreted as more efficient emission for higher dust temperatures. LBGs with lower $M_\star$ are not efficiently detected. The detection is improved compared with the ALMA 850 and 1200 $\micron$ bands for $9<\log (M_\star /\mathrm{M}_{\sun})<9.5$ (Fig.\ \ref{fig:P_M}), particularly at high dust temperatures. Thus, the deep 450 $\micron$ observation is useful to detect high-$T_\mathrm{d}$, intermediate-$M_\star$ objects, which tend to be missed by the current ALMA submm followup of LBGs.
%%The reverse tendency for detection probability may be useful for reveal true evolution by combining two observations.
Some objects with $\log (M_\star /\mathrm{M}_{\sun}) <9$ are also detected. Therefore, in the future, we can obtain more information about low stellar mass objects,
and the future 450 $\micron$ observations we considered here could be even deeper than the current ALMA 850 and 1200 $\micron$ observations in terms of the detected stellar mass range of LBGs.
%%including those missed by the current ALMA observations.

%%subsection{examples of the correction for the temperature bias.}
%%\textbf{Give some simple examples of the correction for the temperature bias for a stellar mass selected sample and a $L_\mathrm{IR}$ selected sample. Maybe we can assume some distribution by a Gaussian centred at $\sim 50$ K, and see how the corrected distribution changes relative to the original distribution. If the change is small, we can judge that the observed statistics of dust temperature is unbiased. (For $\log M_\star /\mathrm{M}_{\sun}=9$--10 with various values of $\mathcal{D}_\star$. If the bias depends on $\mathcal{D}_\star$, it is also an important conclusion.}

\section{Conclusion}\label{sec:conclusion}
We investigate if the current submm observations (represented by ALMA) are fairly tracing the dust temperature in high-redshift ($z\geq 5$) LBGs. To this goal, we perform a simple test using random realizations of LBGs with various stellar masses ($M_\star$), dust temperatures ($T_\mathrm{d}$), and dust-to-stellar mass ratios ($\mathcal{D}_\star$). The values of these three quantities are chosen to cover the parameter space of observed LBGs at $z\geq 5$. We assume that the UV luminosity is strongly correlated with the stellar mass and that the stellar radiation is not highly obscured ($\mathrm{IRX}<10$).
%%We find that high-$T_\mathrm{d}$ objects ($\ge$45 K) are allowed against the physical constrain of IRX, a fact that high-$T_\mathrm{d}$ objects are still the data base for our ALMA detection test. Detection test distinguishes detectable LBGs from 3000 simulated galaxies by ALMA sensitivity values (3$\sigma$) under integration time, 1 hours and 5 hours. 

We find that the dust temperature bias enters differently
depending on the sample selection and the wavelength (in the observer's frame).
If we consider a sample with a fixed range of the total IR dust luminosity $L_\mathrm{IR}$ (IR-referenced sample), the 850 $\mu$m and 1200 $\mu$m samples are biased to low dust temperatures ($\lesssim$45 K) %%at $L_\mathrm{IR}=10^{11}$ L$_\odot$)
even if the IR luminosity is as high as $L_\mathrm{IR}\gtrsim 10^{11}$ L$_{\sun}$.
The 450  $\micron$ band is slightly less biased compared with the longer wavelengths, and is useful to determine the dust temperature; however, it is much shallower. If we select a sample (UV-referenced sample) with a fixed range of the UV luminosity (equivalent to $M_\star$ for LBGs in our model), the dust temperature bias is weaker compared with the IR-referenced sample. Note that the number of detectable LBGs decreases as the redshift becomes higher at 450 $\micron$ since the peak of dust emission SED shifts beyond 450 $\micron$. In contrast, with the negative $K$ correction, detection at 850 and 1200 $\micron$ is not sensitive to the redshift.

Although the dust temperature bias in a UV-referenced sample is milder than that in an IR-referenced one, there are still some biases, which depend on $M_\star$.
There are competing effects between $\mathcal{D}_\star$ and $T_\mathrm{d}$ in determining the bias: Since we exclude high-IRX objects (not typical of LBGs), high-$T_\mathrm{d}$ objects tend to have low $\mathcal{D}_\star$. Thus, more efficient dust emission with a higher $T_\mathrm{d}$ can be counterbalanced by a lower dust abundance.
Since the current ALMA detections of $z\gtrsim 5$ LBGs predominantly sample LBGs with $M_\star >10^{9.5}$~M$_{\sun}$ in the 1200~$\micron$ band, they should not be biased for high $T_\mathrm{d}$ according to our results. However, the detected LBGs have broadly high dust temperatures. Thus, the high $T_\mathrm{d}$ in high redshift LBGs reflects a real trend, not caused by a bias.
We also find that the low detection rates of $z\gtrsim 5$ LGBs with ALMA are also consistent with our results.

The 450~$\micron$ band is differently biased for $T_\mathrm{d}$ from 850 and 1200~$\micron$,
so that it can be useful to obtain an unbiased view of the dust temperature;
however, it is much shallower than the longer-wavelength bands. Thus,
we investigate a possibility of future deep 450 $\micron$ single-dish surveys.
We particularly consider the future 30-m Antarctic THz telescope at New Dome Fuji, of which the low water vapour atmospheric condition improves the sensitivity at 450 $\micron$. Applying the same model but a deeper detection limit, we show that it is possible to obtain an almost complete (i.e.\ without $T_\mathrm{d}$ bias) IR-selected sample in a luminosity range of $L_\mathrm{IR}>10^{11}$ L$_{\sun}$ at $z=5$--7 LBGs and for $L_\mathrm{IR}>10^{11.5}$ L$_{\sun}$ at $z=10$. A UV-referenced sample to be detected by this future telescope shows a bias towards high $T_\mathrm{d}$, but the detection in the intermediate-$M_\star$ range [$9<\log (M_\star /M_{\sun})<9.5$] is much improved compared with the ALMA 850 and 1200 $\micron$ bands, especially at high $T_\mathrm{d}$. Thus, the deep 450 $\micron$ survey is useful to detect high-$T_\mathrm{d}$, intermediate-$M_\star$ LBGs, which tend to be missed in the current submm follow-ups of LBGs.
Besides, the future sensitive telescope is able to detect some LBGs with low IR luminosity [$\log(L_\mathrm{IR}/\mathrm{L}_{\sun})<11$] and low-stellar mass [$\log (M_\star /\mathrm{M}_{\sun}) <9$] at 450 $\micron$. Therefore, to obtain an unbiased and deeper view of the first dust enrichment in the Universe, a future large single-dish telescope capable of observing at short submm wavelengths is useful.

\section*{Acknowledgements}
 
We are grateful to C.-Y. Lin and the anonymous referee for useful discussions and comments.
HH thanks the Ministry of Science and Technology for support through grant
MOST 107-2923-M-001-003-MY3 (RFBR 18-52-52006) and MOST 108-2112-M-001-007-MY3, and the Academia Sinica
for Investigator Award AS-IA-109-M02.

\section*{Data Availability}

Data related to this publication and its figures are available on request from the corresponding author.

%%%%%%%%%%%%%%%%%%%% REFERENCES %%%%%%%%%%%%%%%%%%

\bibliographystyle{mnras}
\bibliography{reference}

% Don't change these lines
\bsp	% typesetting comment
\label{lastpage}
\end{document}